\documentstyle[12pt]{article}
\def\beq{\begin{equation}}
\def\eeq{\end{equation}}
\def\bea{\begin{eqnarray}}
\def\eea{\end{eqnarray}}

\def\ba{\begin{array}}
\def\ea{\end{array}}

\def\,{\"{U}}
\def\6{\.{I}}
\begin{document}

\title{Exact Solutions of the Schr\"{o}dinger Equation with position-dependent effective mass via general
point canonical transformation }
%\author{Cevdet Tezcan, Ramazan Sever\thanks{
\author{Cevdet Tezcan\\
Faculty of Engineering, Ba\c{s}kent University, Bagl{\i}ca Campus, Ankara, Turkey\\[.5cm]
 Ramazan Sever$\thanks{Corresponding author:
sever@metu.edu.tr}$\\
Department of Physics, Middle East Technical University, 06531
Ankara, Turkey }

\date{\today}
\maketitle

\begin{abstract}
\noindent Exact solutions of the Schr\"{o}dinger equation are
obtained  for the Rosen-Morse and Scarf potentials with the
position-dependent effective mass by appliying a general point
canonical transformation. The general form of the point canonical
transformation is introduced by using a free parameter. Two
different forms of mass distributions are used. A set of the energy
eigenvalues of the bound states and corresponding wave functions for
target potentials are obtained as a function of the free parameter.\\

%{PACS:05.20.-y; 05.30.-d; 05.70. Ce; 03.65.-w}\\
%{\it Keywords}{\small : Supersymmetric quantum mechanics, Morse
%potential, Hulth\'{e}n potential, P\"{o}schl-Teller potential,
%Kratzer-like potential, Hamiltonian Hierarchy Method}
\end{abstract}

\baselineskip 0.9cm

\newpage

\section{Introduction}

\noindent Exact solutions of the effective mass Schr\"{o}dinger
equations for some physical potentials have much attention. They
have found important applications in the fields of material science
and condensed matter physics such as semiconductors [1], quantum
well and quantum dots [2], $^{3}H$ clusters [3], quantum liquids
[4], graded alloys and semiconductor heterostructures [5,6].
Recently, number of exact solutions on these topics increased
[6-23]. Various methods are used in the calculations. The point
canonical transformations (PCT) is one of these methods providing
exact solution of energy eigenvalues and corresponding
eigenfunctions [24-27]. It is also used for solving the
Schr\"{o}dinger equation with position-dependent effective mass for
some potentials [8-13]. In the present work, we solve two different
potentials with the two mass distributions. The point canonical
transformation is taken in the more general form introducing a free
parameter. This general form of the transformation will provide us a
set of solutions for different values of the free parameter.

The contents of the paper is as follows. In section 2, we present
briefly the solution of the Schr\"{o}dinger by using point canonical
transformation. In section 3, we introduce some applications for
specific potentials. Results are discussed in section 4.

\section{Method}
\noindent To introduce a general form of PCT with a free parameter,
we start from a time independent Schr\"{o}dinger equation for a
potential $V(y)$

\begin{equation}
\left(-\frac{1}{2}\frac{d^2}{dy^2}+V(y)\right)\phi(y)=E\phi(y)
\end{equation}
\noindent where the atomic unit $\hbar=1$ and the constant mass
$M=1$ are taken. We define a transformation $y\rightarrow x$ for a
mapping $y=f(x)$, we rewrite the wave functions in the form of

\begin{equation}
\phi(y)=m^\alpha(x)\psi(x).
\end{equation}

\noindent Here we assume that mass The transformed Schr\"{o}dinger
equation takes

\begin{eqnarray}
&&\left\{-\frac{1}{2}\frac{d^{2}}{dx^{2}}-\left(\alpha\frac{m^{\prime}}{m}-
\frac{f^{\prime\prime}}{2f^{\prime\prime}}\right)\frac{d}{dx}-\frac{\alpha}{2}\left[\frac{m^{\prime\prime}}{m^{\prime}}
+(\alpha-1)\left(\frac{m^{\prime}}{m}\right)^{2}-\left(\frac{m^{\prime}}{m}\right)
\frac{f^{\prime\prime}}{f^{\prime}}\right]\right.\nonumber\\[0.3cm]
&+&\left.\left(f^{\prime}\right)^{2}~V(f(x))\right\}~\psi(x)=(f^{\prime})^{2}E~\psi(x),
\end{eqnarray}

\noindent where the prime denotes differentiation with respect to
$x$. On the other hand the one dimensional Schr\"{o}dinger equation
with position dependent mass can be written as
\begin{equation}
-\frac{1}{2}\frac{d}{dx}\left[\frac{1}{M(x)}\frac{d\psi(x)}{dx}\right]+\tilde{V}(x)\psi(x)=\tilde{E}\psi(x),
\end{equation}

\noindent where $M(x)=m_{0}~m(x)$, and the dimensionless mass
distribution $m(x)$ is real function. For simplicity, we take
$m_{0}=1$. Thus, Eq. (4) takes the form

\begin{equation}
\left(-\frac{1}{2}\frac{d^2}{dx^2}+\frac{m^\prime}{2m}\frac{d}{dx}+m\tilde{V}(x)\right)\psi(x)=m\tilde{E}\psi(x).
\end{equation}

\noindent Comparing Eqs. (3) and (5), we get the following
identities

\begin{equation}
\frac{f^{\prime\prime}}{2f^\prime}-\alpha\frac{m^\prime}{m}=\frac{m^\prime}{2m}
\end{equation}

\noindent and

\begin{equation}
\tilde{V}(x)-\tilde({E})=\frac{{f^{\prime}}^{2}}{m}\left[V(f(x))-E\right]-
\frac{\alpha}{2m}\left[(\alpha-1)\left(\frac{m^\prime}{m}\right)^2
-\left(\frac{m^\prime}{m}\right)\left(\frac{f^{\prime\prime}}{f^\prime}\right)+
\frac{m^{\prime\prime}}{m}\right]
\end{equation}

\noindent From Eq. (6), we find

\begin{equation}
f^\prime=m^{2\alpha+1}
\end{equation}

\noindent Substituting $f^{\prime}$ into Eq. (7), we obtain the new
potential as

\begin{equation}
\tilde{V}(x)=\frac{{f^{\prime}}^{2}}{m}V(f(x))+(1-m^{4\alpha+1})E-
\frac{\alpha}{2m}\left[\frac{m^{\prime\prime}}{m}-(\alpha+2)\left(\frac{m^{\prime}}{m}\right)^{2}\right].
\end{equation}

\noindent Therefore, the energy eigenvalues and corresponding wave
functions for the potential $V(y)$ as $E_n$ and $\phi_{n}(y)$ become

\begin{eqnarray}
\tilde{E}_n&=&E_n\nonumber\\
\psi_n(x)&=&\frac{1}{m^{\alpha}(x)}\phi_n(y)
\end{eqnarray}
For $\alpha = -1/4$\ Eqs. (9) and (10) reduce to the same form given
in Ref. [13].
\section{Some Applications}
\noindent In this section, we use two different position-dependent
mass distributions. The reference potentials are taken as the
Rosen-Morse [28, 29] and Scarf [30] potentials to get some target
potentials providing us the exact solutions.

\subsection{Mass Distribution $m(x)=a^2/(q+x^2)$}
\noindent The deformed Rosen-Morse and Scarf potentials are

\begin{equation}
V_{RMT}(y)=-V_1~sech_{q}^{2}(\beta y)-V_2~tanh_{q}(\beta y)
\end{equation}

\noindent and

\begin{equation}
V_{ST}(y)=-V_1~sech_{q}^{2}(\beta y)-V_2~sech_{q}(\beta
y)~tanh_{q}(\beta y)
\end{equation}

\noindent where the parameters $V_1$ and $V_2$ are real. The
deformed hyperbolic functions [31] are

\begin{equation}
sinh_{q}y=\frac{e^{y}-q e^{-y}}{2},\quad cosh_{q}y=\frac{e^{y}+q
e^{-y}}{2},\quad tanh_{q}y=\frac{sinh_{q}y}{cosh_{q}y}
\end{equation}
\noindent and
\begin{equation}
cosech_{q}y=\frac{1}{sinh_{q}y},\quad
sech_{q}y=\frac{1}{cosh_{q}y},\quad coth_{q}y=\frac{1}{tanh_{q}y}
\end{equation}

\noindent where $q$ is a real, positive parameter. Recently, it is
found that the deformed hyperbolic potentials can be reduced to the
non-deformed hyperbolic potentials by using a coordinate
translational transformation [32].

\noindent The Schr\"{o}dinger equation with a constant mass for the
Rosen-Morse and Scarf potentials are solved by using
Nikiforov-Uvarov and the function analysis methods [28, 29]. The
exact bound state solutions of the Klein-Gordon and Dirac equations
with equal scalar and vector potentials for these potentials are
also obtained [33, 34].

\noindent We define a parameter $\eta$ to use as a combination of
some parameters

\begin{equation}
\eta=n+\frac{1}{2}-\sqrt{\frac{1}{4}+\frac{V_1}{q\alpha^{2}}}.
\end{equation}

\noindent The energy eigenvalues and corresponding wave functions
for the reference potential $V_{RMT}(y)$ are respectively

\begin{equation}
E_{RMT}(n)=-\frac{V_{2}^{2}}{4\alpha^{2}}\frac{1}{\eta^{2}}-\alpha^{2}\eta^{2}
\end{equation}

\noindent and

\begin{equation}
\phi_{n}(y)=\left(cosh_{q}(\beta
y)\right)^{\eta}\exp\left[-\frac{V_2}{2\beta\eta}y\right]
P_{n}^{-2P_{+},~-2P_{-}}(-tanh_{q}(\beta y))
\end{equation}

\noindent where the quantum number is defined as

\begin{equation}
n=0, 1,
2,\ldots<\sqrt{\frac{1}{4}+\frac{V_1}{q\alpha^{2}}}-\frac{1}{2}.
\end{equation}

\noindent The parameters $P_{+}$ and $P_{-}$ are given by

\begin{equation}
P_\pm=\frac{1}{2}\left[\eta\pm\frac{V_2}{2\beta^{2}}\frac{1}{\eta}\right].
\end{equation}

\noindent Similarly, the energy eigenvalues and the corresponding
wave functions for the reference potential $V_{ST}(y)$ are
respectively
\begin{equation}
E_{ST}(n)=-\beta^{2}\left[n+\frac{1}{2}-\frac{1}{2}\left(\sigma\sqrt{\frac{1}{4}+
\frac{V_1}{q\beta^{2}}+\frac{V_2}{i\beta^{2}q^{1/2}}}+
\tau\sqrt{\frac{1}{4}+\frac{V_1}{q\beta^{2}}
-\frac{V_2}{i\beta^{2}q^{1/2}}}~\right)\right]^{2}
\end{equation}
\noindent and
\begin{eqnarray}
\tilde{\phi}_{n}(y)&=&\frac{1}{[cosh_{q}(\beta y)]^{\omega_{+}
+\omega_{-}}}~exp\left[(\omega_{+}-\omega_{-})~tanh^{-1}(iq^{-1/2}sinh_{q}(\beta
y))\right]\nonumber\\[0.2cm]
&\times&P_{n}^{-2\omega_{+}-\frac{1}{2},~2\omega_{-}-\frac{1}{2}}(iq^{-1/2}sinh_{q}(\beta
y))
\end{eqnarray}

\noindent where, the quantum number is

\begin{equation}
n=0, 1, 2,\ldots<Re \left[\frac{1}{2}\left(\sigma\sqrt{\frac{1}{4}+
\frac{V_1}{q\beta^{2}}+\frac{V_2}{i\beta^{2}q^{1/2}}}+
\tau\sqrt{\frac{1}{4}+\frac{V_1}{q\beta^{2}}
-\frac{V_2}{i\beta^{2}q^{1/2}}}~\right)\right]-\frac{1}{2}
\end{equation}

\noindent The parameters $\omega_{+}$ and $\omega_{-}$ are given by

\begin{equation}
\omega_{+}=-\frac{1}{4}+\frac{\sigma}{2}\sqrt{\frac{1}{4}+
\frac{V_1}{q\beta^{2}}+\frac{V_2}{i\beta^{2}q^{1/2}}},\quad
\omega_{-}=-\frac{1}{4}+\frac{\tau}{2}\sqrt{\frac{1}{4}+
\frac{V_1}{q\beta^{2}}-\frac{V_2}{i\beta^{2}q^{1/2}}}
\end{equation}
\noindent where
\begin{equation}
\sigma=\pm 1\qquad and\qquad \tau=\pm 1.
\end{equation}

\noindent Now, we consider the mass distributions

\begin{equation}
m(x)=\frac{a^{2}}{q+x^{2}}\nonumber\\
\end{equation}

\noindent The mapping function becomes
\begin{eqnarray}
y=f(x)&=&\int m(x)^{2\alpha+1}dx\nonumber\\[0.3cm]
&=&a^{4\alpha+2}\int\frac{dx}{(q+x^{2})^{2\alpha+1}}
\end{eqnarray}

\noindent For $\alpha=-1/4$, Eq. (26) reduces to [13]. Therefore, we
calculate the target potentials for the Rosen-Morse and Scarf
potentials

\begin{eqnarray}
\tilde{V}_{1}(x)&=&m^{4\alpha+1}\left[-V_1~sech_{q}^{2}(\frac{y}{a})
-V_2~tanh_{q}(\frac{y}{a})\right]+(1-m^{4\alpha+1})E_{RMT}(n)\nonumber\\[0.2cm]
                &+&\frac{\alpha}{a^{2}}\left(1+2\alpha\frac{x^{2}}{q+x^{2}}\right)
\end{eqnarray}
\noindent and
\begin{eqnarray}
\tilde{V}_{2}(x)&=&m^{4\alpha+1}\left[-V_1~sech_{q}^{2}(\frac{y}{a})
-V_2~sech_{q}(\frac{y}{a})~tanh_{q}(\frac{y}{a})\right]+
(1-m^{4\alpha+1})E_{ST}(n)\nonumber\\[0.2cm]
               &+&\frac{\alpha}{q^{2}}\left(1+2\alpha\frac{x^{2}}{q+x^{2}}\right)
\end{eqnarray}

\noindent Again for $\alpha=-1/4$, the target potentials reduce to
the ones in Ref. [13]. In order to express the target potentials
$\tilde{V}_{1}$ and $\tilde{V}_{2}$ in terms of $x$, we consider the
following special cases:

\noindent $i)$ For $\alpha=0$
\begin{equation}
y=\frac{a_{2}}{\sqrt{q}}tan^{-1}(\frac{x}{\sqrt{q}})
\end{equation}

\noindent $ii)$ For $\alpha=-1/4$
\begin{equation}
y=a~ln[x+\sqrt{q+x^{2}}]
\end{equation}

\noindent $iii)$ For $\alpha=1$
\begin{equation}
y=q^{-\frac{\tau}{2}}~a^{6}\left[\frac{3}{8}\theta+\frac{3}{8}sin\theta~cos\theta
+\frac{1}{4}sin\theta~cos^{2}\theta\right],
\end{equation}

\noindent where $\theta=tan^{-1}(\frac{x}{\sqrt{q}})$.

\noindent $iv)$ For any $\alpha$.

\noindent Substituting $x=\sqrt{a}~tan\theta$ into the Eq. (26), we
get
\begin{equation}
y=\frac{a^{4\alpha+1}}{q^{2\alpha+1/2}}\int~cos^{4\alpha}\theta~d\theta
\end{equation}

\subsection{Mass Distribution $m(x)=a^2/(b+x^2)^2$}

\noindent We list below the special cases as

\noindent $i)$ For $\alpha=0$

\begin{equation}
y=\frac{a^{2}}{b^{3/2}}\left(\frac{1}{2}sin\theta~cos\theta+\frac{1}{2}\theta\right),
\end{equation}
\noindent where $\theta=tan^{-1}(\frac{x}{\sqrt{b}})$.

\noindent $ii)$ For $\alpha=-1/4$
\begin{equation}
y=a~tan^{-1}(\frac{x}{\sqrt{b}})
\end{equation}

\noindent $iii)$ For $\alpha=1/2$
\begin{equation}
y=\frac{a^{4}}{b^{7/2}}\left[\frac{5}{2}+\frac{5}{16}sin\theta~cos\theta+\frac{5}{24}sin\theta~cos^{2}\theta
+\frac{1}{6}sin\theta~cos^{5}\theta\right],
\end{equation}

\smallskip
\noindent where $\theta=tan^{-1}(\frac{x}{\sqrt{b}})$.

\noindent $iv)$ For any $\alpha$
\begin{equation}
y=a^{4\alpha+2}\int\frac{dx}{(b+x^{2})^{4\alpha+2}}.
\end{equation}

\noindent Substituting $x=\sqrt{b}tan\theta$ into above equation, we
get
\begin{equation}
y=\frac{a^{4\alpha+2}}{b^{4\alpha+2}}\int~cos^{8\alpha+2}\theta~d\theta.
\end{equation}

\subsection{Mass Distribution $m(x)=e^{(-\alpha x)}$}

\noindent From Eq. (26), we get
\begin{eqnarray}
y=f(x)&=&\int~e^{-(2\alpha+1)ax}dx\nonumber\\
      &=&-\frac{1}{(2\alpha+1)a}~e^{-(2\alpha+1)a x}
\end{eqnarray}

\noindent and
\begin{equation}
x=-\frac{1}{(2\alpha+1)a}~ln\left[-(2\alpha+1)ay\right]
\end{equation}

\section{Conclusions}

\noindent We have applied the point canonical transformation in a
general form by introducing a free parameter to solve the
Schr\"{o}dinger equation for the Rosen-Morse and Scarf potentials
with spatially dependent mass. We have obtained a set of exactly
solvable target potentials by using two position-dependent mass
distributions. Energy eigenvalues and corresponding wave functions
for the target potentials are written in the compact form.

\section{Acknowledgements}

This research was partially supported by the Scientific and
Technological Research Council of Turkey.

\end{document}